\documentclass[twocolumn,showpacs,preprintnumbers,amsmath,amssymb]{revtex4}
\usepackage{tabularx,graphicx}\begin{document}
\newcommand{\beq}{\begin{equation}}
\newcommand{\eeq}{\end{equation}}
\newcommand{\beqn}{\begin{eqnarray}}
\newcommand{\eeqn}{\end{eqnarray}}
\newcommand{\bmath}{\begin{subequations}}
\newcommand{\emath}{\end{subequations}}
\title{$E=mc^2, \Delta H = D(H-H)$, and the end of civilization}
\author{J. E. Hirsch }
\address{Department of Physics, University of California, San Diego\\
La Jolla, CA 92093-0319}
 
\date{\today} 
\begin{abstract} 
100 years ago Einstein discovered $E=mc^2$, the secret energy stored in ordinary mass.
$\Delta H = D(H-H)$ is the chemical energy released in chemical bond
formation between two H atoms. The failure to
recognize the enormously different energy scales in those two equations reflected in current
events may start a chain reaction this very year, on the one-hundredths anniversary of 
Einstein's discovery, that leads 
 to the end of civilization. Due to the confluence of a particular set of circumstances, this particular moment is more dangerous than any other in the
 history of nuclear weapons. Physicists have a special responsibility to do their
utmost to prevent this from happening. This paper is a call to arms. A  principle to underpin nuclear non-proliferation
and enhance stability is advocated.

 \end{abstract}
\maketitle
\section{ Einstein's legacy} 
100 years ago, on September 27, 1905, A. Einstein wrote ``It is not impossible that with bodies whose energy-content is variable to a high degree (e.g. with radium salts) the theory may be successfully put to the test'',   in connection with the
 equation $E=mc^2$ that he had just discovered\cite{einstein} .  Today, Einstein is revered by scientists
 as well as the world at large as the greatest scientist of all times, and this full year has been dedicated
  to honor him and
 his momentous contributions. Yet there is an imminent danger
 that this will  also be the year where Einstein's famous equation sets out
 a chain of events  that leads to the destruction of civilization. 
 This chain of events will be triggered
 by the disregard of politicians to the vastly different energy scales involved in
 nuclear and  chemical reactions\cite{weisskopf}, and the failure of scientists to 
 call this fact forcefully enough to theirs and the public's attention.
  
 It is the responsibility of all humans,
 but especially of physicists, to do their utmost to preserve Einstein's legacy and prevent
 developments that could lead survivors of a nuclear holocaust to remember Einstein as
 the greatest criminal of all times. A set of confluent circumstances makes this particular moment
 more dangerous than any other in the history of nuclear weapons. This paper is a call to arms.
 
 \section{ Einstein and nuclear weapons} 

 It took 14 years after Einstein's  discovery of $E=mc^2$ for the world at large to discover
 Einstein. On the morning of  November 7, 1919, newspapers all over the world reported that Einstein's prediction of
 gravitational deflection of light had just been verified in a solar eclipse, making Einstein from then
 on a celebrity throughout the civilized world. 
 
 It would only take  6 years after another action of Einstein for the world to learn about its
 consequences.  On  August 2nd, 1939,  
Einstein  wrote a letter to US President Roosevelt to point out that the newly discovered phenomenon of nuclear fission  ``would also lead to the construction of bombs'',  and urged Roosevelt to support the work that would lead to nuclear bombs. The concerted work of many theoretical and applied physicists
in the ensuing years under the Manhattan project led to the first controlled nuclear 
chain reaction in December  1942,  and shortly thereafter to the first nuclear explosion in
July   1945. The world
woke up on    August 6, 1945, to discover the horrors of a nuclear bomb used in an act of war.
The full extent of the devastation caused by the nuclear explosions in Hiroshima and Nagasaki
would only become clear several years later, as the long term effects of the unleashed
radioactivity were making their way through the molecules, cells and tissue of the human
bodies that had been exposed.

Einstein himself was of course fully conscious of the enormous responsibility on his shoulders, and in fact in the years after his fateful letter to Roosevelt he opposed the use of nuclear bombs in World War II.  After the
war, he devoted large efforts to work for international cooperation and spoke out fervently in favor of
nuclear disarmament. In Einstein's own  words: {\it ''I believe America may totally succumb to the fearful militarisation which engulfed Germany at the beginning of the 20th century."}, and 
{\it "In all countries power lies in the hands of ambitious power-hungry men. This is true whether the political system is dictatorial or democratic. Power relies not only on coercion, but on subtle persuasion and deception through the educational system and the media of public information."} These words ring fully true today.
 
\section{Weapons of mass destruction}

The term 'weapons of mass destruction (WMD's) is used to describe chemical, biological, radiological and nuclear weapons. Here we will focus on chemical
weapons and nuclear weapons because they present the starkest contrast, 
and discuss what is the significance that they are lumped into the
common concept of WMD in the formulation of policy.

Chemical weapons were first used on a wide scale in World War I. Chlorine gas was first used
by Germany against French forces on April 1915, and by British against German forces
shortly thereafter\cite{ww1}. Later, phosgene and mustard gas were used. Protection methods against
poison gases were however quickly developed, and proved highly effective. The total number
of deaths from chemical weapons in WWI is estimated to be around 90,000\cite{ww1}, approximately
$1\%$ of the total number of casualties on the battlefield. Thus, $100$-times more 'mass destruction'
was caused by conventional weapons than by chemical weapons in a conflict where 
chemical weapons were used almost from the beginning. As scientists that deal with
objective reality we should object to the label 
'weapon of mass destruction' being applied to weapons that cause 100 times fewer casualties than
weapons of non-mass-destruction, unless all weapons are made to   fall under such category.

The term 'weapons of mass destruction (WMD's) has become a central element in the formulation
of United States policy in recent years, especially since Iraq's invasion of Kuwait in 1990. The US invasion
of Iraq in March 2003 was launched with the stated purpose of ridding Iraq of chemical and possibly
biological weapons that Iraq was known to possess before 1990. At the same time the point was
made that Iraq one day would possess nuclear weapons, which would heighten the dangers more. This could appear to be a logical inference if there was a logical link between different WMD's, in particular 
chemical and nuclear weapons. There is however no scientific nor technological link between
chemical weapons and nuclear weapons, nor as discussed above is there any relation between their destructive power that would not also include conventional weapons. Consequently there should not
be a link in policy decisions or statements, unless one accepts a premise that policy can be disconnected
from reality. Scientists should be at the forefront in exposing the flaw of such statements, but their voices
have not been heard.

\section{physics and chemistry}

Chemical weapons involve chemical
 reactions, nuclear weapons involve nuclear fission or fusion reactions. The energy stored in
 the H-H  chemical bond is of order $5eV$, while the energy stored in a H atom through 
 $E=mc^2$ is of order
 $10^9$ eV. Even though only a tiny fraction of the mass of the atoms involved  is released as energy
 in a nuclear fission
 reaction, the fission reaction is over a million times
 more powerful than the most potent chemical reaction\cite{weisskopf}, and even higher for a fusion reaction.

 The detailed processes by which chemical weapons and nuclear weapons cause harm
 to human tissue and ultimately death need not concern us here. The single scaling factor
 of 1 million is all that matters to set the scale of destruction achieved by these weapons,
 since destruction involves energy transfers. From this it follows that if chemical weapons
 can kill 1000 humans, nuclear weapons can kill 1 billion humans. Indeed existing
 nuclear arsenals can destruct humanity many times over.
 
 In World War I, the initial use of chemical weapons immediately led to retaliation by the other
 side with more potent chemical weapons, and their use continued  to escalate by
 both sides\cite{ww1}. The cycle was stopped not because human
 rationality prevailed, rather because of technological reasons, namely that protection against
 chemical weapons was quickly developed. We scientists know for a fact that protection
 against large scale nuclear weapons use cannot be developed. 
 
  The cycle of nuclear weapons use in World War II did not escalate only because at that time
  only  the United States possessed  nuclear weapons. That is no longer true today.
 We know that human
 nature has not radically changed since World War I, so that an escalating cycle of nuclear
 attack and retaliation is likely to continue until total destruction.
  
 We scientists know these facts, the general population is not as fully aware of them. It is our duty as well
 as in our own interest to insert rationality in debates and policy decisions on 
 nuclear weapons. Our professional organizations have not been  vocal on these issues, that needs to change.

\section{physicists' responsibility}

As physicists we bear a special responsibility, because a physicist discovered $E=mc^2$, because
physicists discovered nuclear fission and fusion reactions, because physicists were the main actors
in the Manhattan project, and because of physicists deep involvement in the 
development of nuclear weapons ever since. As physicists we know better than other scientists
and the general public the awesome destructive power of nuclear weapons, and we have 
a responsibility to share that knowledge with others.

Physicists that lived through the Manhattan project years  felt that responsibility deeply, and many prominent physicists were active
individually and through several organizations such as The Bulletin of the Atomic Scientist, Emergency Committee of 
Atomic Scientists, Pugwash, Federation of Atomic Scientists,
contributing their efforts to 
prevent  nuclear arms buildup and related dangerous policies.

However this is no longer happening to the same extent  today. Even though many of these organizations
still exist, their structures have aged and many of their founding members have retired. The younger
generation of physicists has not felt the same call to become involved in nuclear weapons policy issues,
because the reality of nuclear weapons use is not part of their personal experience.
There have not been
strong statements from the arms control organizations nor professional scientific societies deploring
the characterization of nuclear and chemical weapons under the same label of 'WMD".

We need to make a renewed effort to feel the urge that the older generation felt to
combat the potential destructive use of nuclear weapons, otherwise we may be engulfed in
events that once started will be impossible to stop. In Einstein's powerful words,
{\it "In our time, scientists and engineers carry a particularly heavy burden of moral responsibility, because the development of military means of mass destruction is dependent on their work."}

 True, but let the chemists worry about chemical weapons and  biologists about biological weapons; physicists should
 focus and continue taking the lead in worrying about the nuclear weapons that their profession created!

\section{chain reactions}
Einstein's 1905 paper set off a chain reaction that led in a series of many steps
to the nuclear arsenals of today. Chain reaction is what leads in a nuclear device from ignition where a few Uranium 235 atoms split to the full explosion. Once a chain reaction is set into motion it is very difficult or impossible to stop. That is why it is essential to stop a chain reaction before it starts or at its very early stages.  

I do not see a point where the chain reaction that led from 1905 to today could have been stopped. 
Once $E=mc^2$ was revealed, the discovery of fision and fusion had to happen, and there has never
been a time in history where scientific progress has not been exploited for military application. Even if
WW2 had ended before Hiroshima, a nuclear device wold have been used in a military
action eventually to test its awesome power, and humanity was fortunate that this hapened when 
only one nation
had achieved that technology. However I argue here that a specific imminent event 
will trigger a chain reaction
that will lead with very high probability to the destruction of civilization: 
{\it a new use of a nuclear bomb by a nuclear nation against
a non-nuclear one}. I furthermore argue that we are still at a point where this chain of events can be
stopped.

Some may argue  that the point of no return has already been crossed, when the first nuclear weapons were created or used, and that their
large scale use is inevitable in the long run. While impossible to prove wrong, this may also be a self-fulfilling prophecy.
After the horrors of Hiroshima and Nagasaki the world has relied on a universally agreed-upon taboo against the use of all nuclear weapons, no matter how small,  and on the Nuclear Non-Proliferation Treaty (NPT). 
Nations have been encouraged to abandon nuclear weapons ambitions under the promise of a 
general effort towards nuclear disarmament. Once that taboo is broken the trend will
be  reversed, and nations will race to acquire nuclear weapons for protection,
deterrence  and retaliation.
Any regional conflict will have a high  potential to explode into all-out nuclear war, with untold consequences.

\section{The way to stop it}
There is no rational reason why nonnuclear nations will choose to forego the development of nuclear weapons technology if a nuclear nation today uses a nuclear bomb against a non-nuclear adversary, other 
than the fear to be attacked before it reaches the goal. Once a nation acquires a few nuclear bombs,
it possesses a powerful deterrent against nuclear attack even by a far superior adversary:
the fact that it can retaliate even with one hit. Hence even if all nations had nuclear weapons
today it would be a relatively more stable situation. Of course the probability that an unidentifiable
terrorist group acquires and uses a nuclear weapon increases, but nuclear nations no matter
how 'rogue' have every incentive to avoid such an event. If a nuclear nation nukes a non-nuclear
nation however, terrorist groups  that are sympathetic to the victim nation
will have infinitely more incentive to find a way to retaliate 
in the same kind, and eventually will succeed.

It is also inconceivable that the United States and other nuclear countries today will be able to stop the
182 non-nuclear nations from developing nuclear weapons solely by provoking  fear of military action.
The NPT relies on voluntary compliance  by nations following a rational goal, and  will become
untenable if the goal is no longer rational because it leaves the nation exposed to a nuclear
attack by the nuclear nations. Such reasoning is likely to be used even by nations that
today are friendly to the United States and other nuclear nations.

Even in the current situation, the NPT is not providing a strong incentive for non-proliferation, because nuclear
nations are dragging their feet in fulfilling their promises of arms reduction. Instead, a much more stable equilibrium would be achieved if all nuclear nations pledged today 
never to use a nuclear weapon against a non-nuclear adversary (by non-nuclear adversary it is meant
an adversary that does not possess nuclear weapons, even if it is in full control of civilian
nuclear technology). Similar proposals, such as "no first use" pledge\cite{nofirstuse} have been discussed
in the past. The pledge discussed here is more restricted than 'no first use' since it would allow
nuclear nations to strike first at other nuclear nations. 

It is difficult to imagine
sequences of events where such a pledge would leave a nuclear nation in 
a situation where it would need to break the pledge for its survival or even to gain significant  advantages.
If nuclear nations made such a pledge and planned their military policy with the intent
of abiding by the pledge, they could make sure that their other military resources allow them
adequate protection and even adequate offensive capability against non-nuclear 
countries  that possess any other weapon capability, including other "WMD's". I 
 realize that powerful nation may have geopolitical goals that would require use of 
military force in an offensive way\cite{rice}. However this does not have to
involve nuclear weapons. 

Such a pledge by nuclear nations would provide a  powerful incentive for non-nuclear nations to not develop
nuclear weapons, providing them with a powerful shield in the form of the assurance that no nation would ever
use nuclear weapons against them.
Instead, nuclear nations would not enjoy such a privileged status, which in turn would provide a strong  incentive for 
nuclear nations with small arsenals to voluntarily disarm. Non-nuclear nations would become nuclear at their own risk.

By the same token,  in the absence of such a pledge, if military policy and planning of a nation relies on the
use of nuclear weapons against non-nuclear adversaries, events will eventually lead
that  nation to a situation where it will become essentially impossible $not$  to use a nuclear
weapon against a non-nuclear country. I  argue that we are
almost at that point today.

\section{The United States nuclear weapon policy}

Nuclear weapons policy in the United States has steadily evolved in recent years
 towards
integration of conventional and nuclear forces for military operations in restricted theaters.
The United States has never renounced first use of nuclear weapons against conventional
forces, citing as example a Soviet invasion of Western Europe. However recent 
changes indicate that the United States is now prepared to use nuclear weapons,
even in a pre-emptive way,
against enemies that do not possess nuclear weapons and do not represent a threat to
the survival of the United States nor of any ally of the United States.

The document "Nuclear Posture Review"\cite{posture} from the Department of Defense was
 delivered to Congress in December 2001 and represents official United States policy.
 Not all of its contents have been made public, but the ones that have reveal 
 the essence of the new US nuclear weapons policy. It envisages a "new mix"
 of nuclear and non-nuclear capabilities for a "diverse set of potential adversaries", to provide 
 "flexible, pre-planned non-nuclear and nuclear options". For example, it
 states  "Composed of both non-nuclear systems and nuclear weapons, the strike element of the New Triad can provide greater flexibility in the design and conduct of military campaigns to defeat opponents decisively. Non-nuclear strike capabilities may be particularly useful to limit collateral damage and conflict escalation. Nuclear weapons could be employed against targets able to withstand non-nuclear attack, (for example, deep underground bunkers or bio-weapon facilities)."
 Classified presidential directives  spell out further the policies contained in that document\cite{christensen}.

 A draft document  "Doctrine for Joint Nuclear Operations"
 from the U.S. Joint Chiefs of Staff has been made public
 and is available on the world wide web, one version dated September 2003\cite{doctrineold} and a
revised one dated March 2005\cite{doctrine,christensen}. This document is not yet official US policy but it
 is reported that its adoption is imminent\cite{wash}. The document 
 describes in chilling detail scenarios under which nuclear weapons will be used
 against non-nuclear nations. Even if some changes are made before final adoption,
 because  the document   reflects the essentials of the policy contained in
 "Nuclear Posture Review" such changes will only be cosmetic (there is no substantial difference between the
 2003 and 2005 versions in the issues of concern here). The scenarios described
 apply $literally$ to events about to unfold.
 
 In the following I describe a series of events that will lead with near certainty
  in the immediate future
 to implementation of this new nuclear policy and use of a nuclear weapon against a
 non-nuclear nation: Iran.
 
\section{Iran: the imminent danger}

A tense situation has developed with Iran due to the stated desire of Iran to implement
the uranium enrichment cycle for use in nuclear reactors. The US opposes this due to stated fears
that Iran would eventually divert material to build nuclear weapons. It has put pressure on European nations members of the IAEA (International Atomic Energy Agency) 
to refer Iran to the UN Security Council (SC)  for the
consideration of sanctions. Russia and China have explicitly stated that they would veto any SC
resolution to impose sanctions on Iran, as they consider Iran's desire to be legitimate 
and allowed under the Nuclear Non-Proliferation treaty of which Iran is signatory. 
When the issue reaches the Security Council and a sanctions resolution is vetoed, the US will
be left with no diplomatic options, only military ones. 

The US has not engaged in direct negotiations
with Iran and other concerned nations to attempt to reach a mutually acceptable agreement.
The fact that it has pushed for the IAEA resolution to refer Iran to the Security Council on the face of explicit opposition by China and Russia is puzzling, since such referral will not put additional pressure
on Iran, given the promised veto of those two SC members. It ceases to be puzzling however  if
 a US decision to resort to 
a military option has already been made and the purpose of diplomacy is only
to provide cover for planned military action. US President George Bush has explicitly refused to
take the military option off the table\cite{christian}. The only realistic military option for the US is aerial bombing
of Iranian installations, since a ground attack would likely be met with fierce Iranian resistance
that an overstretched US military would not be able to overcome.

\section{Sequence of events to come}
Assuming Iran is referred to the UN Security Council and no sanctions are imposed, US military
action appears unavoidable. Such   action is likely to include the use of low-yield nuclear weapons, for the reasons discussed in what follows.

Iran has signed and ratified the Chemical Weapons Convention (in 1993 and 1997 respectively) that requires it to terminate production and eliminate stockpiles over a period of years, however it is likely to still have supplies\cite{iranchem,iranchem2,cia}.  Proof that Iran still posesses some chemical weapons should not be difficult for the Bush administration to present, together with arguments that Iran has used chemical weapons in the Iran-Iraq war (conveniently omitting the fact that it was responding to chemical attacks by Iraq) and that it could use them again against US and British forces in Iraq. The Pentagon document "Doctrine for Joint Nuclear Operations"\cite{doctrine}
explicitly states that "Geographic combatant commanders may request Presidential
approval for use of nuclear weapons for a variety of purposes that include  "To demonstrate US intent and capability to use nuclear weapons to deter adversary use of WMD". 

If only conventional bombs are used in an unprovoked US attack on Iran, Iran is likely  to retaliate with a barrage of missiles against US and British forces in Iraq, and possibly Israel, as well as possibly a ground invasion of Iraq, that the 150,000 US troops in Iraq would not be able to withstand. Iranian missiles could potentially include chemical warheads, and it certainly would be impossible to rule out such possibility. 
The Pentagon document "Doctrine for Joint Nuclear Operations"\cite{doctrine} explicitly states that nuclear weapons
may be used by the US in the following situations:  "Against an adversary using or
 intending to use WMD against US, multinational or alliance
forces or civilian populations",  "To counter potentially overwhelming adversary conventional forces
including mobile and area target (troop concentration)", "Attacks on adversary installations including
WMD, deep, hardened bunkers containing chemical or biological weapons or the C2 infrastructure required
for the adversary to execute a WMD attack against the United States or its friends and allies", "For rapid
and favorable war termination on US terms", "To ensure success of US and multinational 
operations". Each of these scenarios applies $literally$ to the Iran situation under discussion.

Early use by the US of low-yield nuclear bombs with equal or better bunker-busting ability 
than conventional bombs targeting Iranian nuclear, chemical and missile installations
would be fully rational
under the conditions described above, and would have the added benefit of sending a clear message
to Iran that any response would be answered by an immensely more devastating nuclear attack.
The US Senate is likely to have already approved the deployment of tactical nuclear weapons in the Persian Gulf, given that Iran has missiles that it can equip with chemical warheads at a moments notice. Such a grave measure could be carried out without public disclosure, with the argument that disclosing it would endanger coalition troops in Iraq. Once military action starts the US Senate and a large part of  the American public 
are likely to support the use of low yield nuclear weapons
to destroy Iranian installations, since this would protect  the lives 
of 150,000 US soldiers  that  would otherwise be at great  risk.

\section{summary}
The points discussed in this paper have been separately discussed by other authors in different contexts. 
Still I believe that  given the current situation it is of extreme urgency to
consider the totality of arguments given in this paper, that lead either to a rational course of 
action or an irrational one. I argue that never before, including the Cuban missile crisis, have we been so close to a point where
a probable sequence of events about to unfold would lead to the destruction of humanity. Unlike the Cuban missile
crisis, if this chain of events gets started this time it is not likely to lead to total destruction in a matter of
days or months; nevertheless we will have entered a different world, and even if the time scale of retaliation
and escalation is years rather than days, the chain reaction that will follow  will be unstoppable. Furthermore,
this moment is far more dangerous than the Cuban missile crisis because there is no immediate deterrent
to the initial event that will get this chain reaction started.

We have discussed in this paper the following points: (i) Chemical weapons are qualitatively different
from nuclear weapons. Nuclear weapons are in a class by themselves and it is extremely  dangerous to lump them into the same category
with other so-called 'WMD's"; 
(ii) A country using a nuclear weapon against a non-nuclear country will start a chain of events  that
will lead unavoidably to the destruction of civilization; (iii) Physicists have a special responsibility
to care about these issues and devote effort to help the public understand them and to influence
government policy;
(iv) The situation with Iran is extremely close to a point of no return, beyond which the use
of nuclear weapons will become unavoidable; (v) A rational strategy does exist to create a
substantially more 
stable situation regarding nuclear weapons and provide a real incentive for non-proliferation,  that nuclear countries pledge to under no
circumstances use a nuclear weapon against a non-nuclear adversary.

\section{A call to arms}
As argued above, physicists have an enormous responsibility on their shoulders as the creators of
nuclear weapons.
A petition by physicists opposing pre-emptive nuclear strikes against non-nuclear countries
 is currently on the world wide web\cite{petition}, URL http://physics.ucsd.edu/petition/.
All physicists should consider signing this petition.

In connection with the dangerous Iran situation, US physicists should make their voices be heard urgently and loudly,
communicating with Congresspeople and leading figures which can influence government policy, and writing letters
to local and national media demanding full disclosure of government actions and plans regarding the use
of nuclear weapons, retraction of the policies described  in the documents
"Nuclear Posture Review" and  "Doctrine for Joint Nuclear Operations", 
and stating vehemently their opposition to any course of action that would lead to the necessity of using nuclear weapons  
imminently against any non-nuclear nation and in particular  Iran. European physicists should  likewise  attempt to influence the upcoming actions of the "European partners"
in the IAEA to avoid further escalation of the Iran situation.

Assuming the Iran question is resolved in the next few months without leading to nuclear war, physicists should
be at the forefront of a world-wide effort, that may extend over several years,
 to get humanity to demand that {\it all governments of nuclear countries  pledge absolute
renunciation of the use of  nuclear weapons
against non-nuclear countries}. Such pledge should be individually made and
 incorporated in the law and the constitution of each nuclear country,
and it should not be put as condition by any one country that  all nuclear countries subscribe to the pledge. However,
nuclear countries that do not sign  the pledge would become pariahs in the world community.
{\it Citizens of the world should actively boycott all products and services of nuclear nations that have not signed the pledge.}
It is likely that the United States will be reluctant to sign, however even a country as powerful as the
United States will feel the pressure of citizens of all other countries as well as their own
boycotting all American-made products and services.
Foregoing the enjoyment of American productivity is a small price to pay when the survival
of humanity is at stake.

Let us all take some time out of our everyday scientific
activities to work towards ensuring that we and our descendants will be
around to enjoy the fruits of our labor, relish the benefits that flowed from Einstein's discovery of $E=mc^2$,
and prevent  that the world ever regrets that $E=mc^2$ was discovered.

\acknowledgements 
 I am  grateful to the many devoted people of the arms control community and other concerned people
that are watchful of these developments  and put on the web documents
concerning nuclear arms policies and practices, and to the journalists that call these developments to the public attention.


\begin{references}
\bibitem{einstein} A. Einstein, Ann. Physik {\bf 18}, 639 (1905).
\bibitem{weisskopf} V.F. Weisskopf, http://www.harvardsquarelibrary.org  \newline
/speakout/weisskopf.html.
\bibitem{ww1} http://www.firstworldwar.com/weaponry/gas.htm
\bibitem{nofirstuse} $http://en.wikipedia.org/wiki/No_first_use$
\bibitem{rice} Speech by US Secretary of State, September 30, 2005, \newline
$http://www.state.gov/secretary/rm/2005/54176.htm$
\bibitem{posture} http://www.globalsecurity.org/wmd/library/ \newline policy/dod/npr.htm
 \bibitem{christensen} Hans M. Christensen, $http://www.armscontrol.org/act/2005_09/Kristensen.asp$
\bibitem{doctrineold} "Doctrine for Joint Nuclear Operations", September 2003 version, 
 $http://www.armscontrolwonk.com/?c=nuclear-weapons$
\bibitem{doctrine} "Doctrine for Joint Nuclear Operations", September 2003 version, 
$www.globalsecurity.org/wmd/library/ \newline policy/dod/jp3\_12fc2.pdf$
\bibitem{wash} Walter Pincus, Washington Post, September 11, 2005.
\bibitem{christian} Tom Regan, Christian Science Monitor, August 17, 2005, http://www.csmonitor.com/2005/0817/dailyUpdate.html
\bibitem{iranchem} http://www.globalsecurity.org/wmd/world/iran/cw.htm
\bibitem{iranchem2} http://www.fas.org/nuke/guide/iran/cw/
\bibitem{cia} Unclassified CIA report to Congress, December 31, 2003,\newline
$http://www.cia.gov/cia/reports/721_reports/july_dec2003.htm$
\bibitem{petition} http://physics.ucsd.edu/petition/
\end{references}
\end{document}